\documentclass[a4paper]{article}

\usepackage{INTERSPEECH2020}
\usepackage{todonotes}
\usepackage{multirow}
\usepackage[tone]{tipa}
\usepackage{subcaption}
\usepackage{comment}

\newcommand\mk{\kern-0.2em}
\newcommand\K{\kern0.2em}

\newcommand\blfootnote[1]{%
  \begingroup
  \renewcommand\thefootnote{}\footnote{#1}%
  \addtocounter{footnote}{-1}%
  \endgroup
}

\title{That Sounds Familiar: an Analysis of Phonetic Representations Transfer Across Languages}
\name{Piotr \.{Z}elasko$^1$, Laureano Moro-Vel\'{a}zquez$^1$, Mark Hasegawa-Johnson$^3$, Odette Scharenborg$^4$, Najim Dehak$^{1,2}$}
\address{
  $^1$Center for Language and Speech Processing,
  $^2$Human Language Technology Center of Excellence, \\ Johns Hopkins University, Baltimore, MD, USA \\
  $^3$ECE Department and Beckman Institute, University of Illinois, Urbana-Champaign, USA \\
  $^4$Multimedia Computing Group, Delft University of Technology, Delft, the Netherlands
 }
\email{piotr.andrzej.zelasko@gmail.com}

\begin{document}

\maketitle
\begin{abstract}
    Only a handful of the world's languages are abundant with the resources that enable practical applications of speech processing technologies. 
    One of the methods to overcome this problem is to use the resources existing in other languages to train a multilingual automatic speech recognition (ASR) model, which, intuitively, should learn some universal phonetic representations.
    In this work, we focus on gaining a deeper understanding of how general these representations might be, and how individual phones are getting improved in a multilingual setting.
    To that end, we select a phonetically diverse set of languages, and perform a series of monolingual, multilingual and crosslingual (zero-shot) experiments.
    The ASR is trained to recognize the International Phonetic Alphabet (IPA) token sequences.
    We observe significant improvements across all languages in the multilingual setting, and stark degradation in the crosslingual setting, where the model, among other errors, considers Javanese as a tone language.
    Notably, as little as 10 hours of the target language training data tremendously reduces ASR error rates.
    Our analysis uncovered that even the phones that are unique to a single language can benefit greatly from adding training data from other languages - an encouraging result for the low-resource speech community.
\end{abstract}

\section{Introduction}
\label{sec:introduction}

\blfootnote{This work was funded by NSF IIS 19-10319.  All findings and conclusions are those of the authors, and are not endorsed by the NSF.}
\blfootnote{We intend to release the experimental code after publication.}
Automatic speech recognition (ASR) is one of the most impactful technologies that have been deployed on a massive scale with the beginning of the 21st century. It enabled the ubiquitous appearance of digital assistants, which help many of us with everyday tasks. Combined with spoken language understanding (SLU), ASR has the potential to accelerate, scale, and even automate numerous processes that require communication, such as requests for support, or inquiries for knowledge. Regrettably, only a small number of the world's languages - mostly the ones that became widely used as a result of colonialism~\cite{samarin1984linguistic,helgerson1998language,pennycook2002english} - are sufficiently resourced to build speech processing systems.  In the world today, people without the resources to become literate in one of the well-resourced languages are thereby placed on the wrong side of the digital divide~\cite{Morrell2020}; if ASR depends on huge investments in speech resources, it does nothing to help them. Our study aims to provide a small step forward in challenging the \emph{status quo}.

Past research~\cite{schultz1998multilingual,loof2009cross,Swietojansk2012unsupervised} addressed this problem, finding that existing resources for other languages can be leveraged to pre-train, or bootstrap, an acoustic model, and then adapt it to the target language, given a small quantity of adaptation data.

For instance, some authors have used the IARPA Babel project corpora to train multilingual ASR~\cite{knill2013investigation} in the task of spoken term detection in under-resourced languages. 
More recent studies employ up to 11 different well-resourced languages to create a nearly-universal phone encoder that would be suitable for almost any language, followed by a language dependent allophone layer and a loss function independent for each language \cite{li2020universal}. This solution provides up to 17\% Phone Error Rate (PER) improvement in the under-resourced languages studied. Fine-tuning a pre-trained multilingual ASR model has been shown to reduce the dataset size requirements when building ASR for a new target language~\cite{dalmia2018sequence}.

Given these past findings of the multilingual ASR community, it is clear that leveraging the combined resources leads to improvements in transcription for less-resourced languages, compared to using the scarce monolingual resources alone. Yet, besides reporting the phone error rate (PER) or word error rate (WER) improvements, the past studies have not comprehensively explored what the model is learning. We know that the speech models learn phonetic representations~\cite{nagamine2015exploring} -- however, which phones' representations are the most amenable to crosslingual transfer? Are there some specific phonetic categories that can be improved more than others? Are we able to improve the representation of phones that are not observed outside of the target language? Are there phones whose models are universal? These are some of the questions we address.

To that end, we train an end-to-end (E2E) phone-level ASR model with recordings in 13 languages, vastly differing in their phonetic properties. We obtain the IPA phonetic transcripts by leveraging the LanguageNet grapheme-to-phone (G2P) models\footnote{https://github.com/uiuc-sst/g2ps}.
The model predicts sequences of phonetic tokens, meaning that each phone is split into basic IPA symbols. We conduct three types of experiments: monolingual, as a baseline; multilingual, to investigate the effects of pooling all languages together; and crosslingual, or zero-shot, to analyze which phone representations are general enough for recognition in an unseen, and possibly even an unwritten, language. These experiments are followed by an analysis of errors -- at a high-level, between languages and experimental scenarios, and at a low-level, considering individual phones, as well as manner and place of articulation.

\section{Experimental setup}
\label{sec:experimental_setup}

The languages used in this study were chosen for the diversity of their phoneme inventories (see Table~\ref{tab:dataset_hours}): 
Zulu has clicks, Bengali was chosen as a representative language with voiced aspirated stops, Javanese for its slack-voiced (breathy-voiced) stops, and Mandarin, Cantonese, Vietnamese, Lao, Thai, and Zulu are tone languages with very different tone inventories.  Table~\ref{tab:dataset_hours} reports the number of distinct IPA symbols used to code the phoneme inventories of each language, e.g., the five Amharic phonemes~\textipa{/t/},\textipa{/t'/},\textipa{/S/},\textipa{/tS/},\textipa{/tS'/} are coded using only three IPA symbols:~\textipa{[t]},~\textipa{[S]}, and~\textipa{[']}. French, Spanish, and Czech have the most well-trained G2P transducers --- each is trained using a lexicon with hundreds of thousands of words.

Speech data for these languages were obtained from the GlobalPhone~\cite{schultz2002globalphone} and IARPA Babel corpora. GlobalPhone has a small number of recording hours for each language - about 10 to 25h - and represents a simpler scenario with limited noise and reverberation. On the other hand, Babel languages have more training data, ranging from 40 to 126 hours, but these are significantly more challenging due to more naturalistic recording conditions. We use standard train, development, and evaluation splits for Babel. We do the same with GlobalPhone languages whenever the documentation provides standard split information - otherwise, we chose the first 20 speakers' recordings for development and evaluation sets (10 speakers each), and train on the rest.

We use an end-to-end\footnote{
We do not claim that an E2E system is necessarily better or worse for this task than an LFMMI-trained hybrid system. We intend to compare the performance with a hybrid system in the future.
} 
(E2E) automatic speech recognition system based on combined filterbank and pitch features~\cite{ghahremani2014pitch}, trained with joint CTC and attention objectives~\cite{kim2017joint}, leveraging the ESPnet toolkit~\cite{watanabe2018espnet}. The setup is based on~\cite{watanabe2017language}, however, instead of using a CNN-BLSTM encoder and LSTM decoder, we use the same Transformer components and configuration as~\cite{karita2019comparative}. To accommodate its $O(n^2)$ memory requirements, we discard all utterances which exceed 30 seconds of speech (about 1.7\% of utterances). An identical architecture is used for each experiment.

Notably, the main difference, compared to~\cite{karita2019comparative}, is that our system is trained to predict IPA~\cite{international1999handbook} symbols instead of characters. The output layer consists of IPA symbols that are present in the training data for a given experiment. Modifier symbols, such as long vowels \textipa{[:]} or high tone levels \textipa{[\tone{55}]}, are also separate symbols in the final layer. Due to this output layer construction, our model can share more parameters between phones with and without suprasegmental or tonal symbols (e.g. non-tonal \textipa{[a]} and tonal \textipa{[a\tone{55}]}), than would be possible should these phones be represented by two discrete symbols. Also, the model learns the concept of \emph{tonality} or \emph{stress} separately from the phones. For instance, the primary stress symbol \textipa{["]}, which is present in both \textipa{["a]} and \textipa{["i]}, is learned as a separate label from \textipa{[a]} and \textipa{[i]}. The potential downside is that the model theoretically might recognize non-existent phones, such as \textipa{[s\tone{22}]}, however in practice we observe that such non-existent phones are hardly hypothesized. We take this phenomenon into account during the evaluation, where we also treat every modifier symbol as a separate token. That means that if the reference transcript has a vowel with a tone symbol, and the ASR recognized the vowel correctly but omitted the tone, it would be counted as one correct token and one deletion instead of a single substitution. Because of that, we do not necessarily measure phone error rate (PER) - rather, phonetic token error rate (PTER).

\begin{table}[th]
  \caption{Amount of data as measured in hours for training, development and evaluation sets for each language. The data for the first five languages is from GlobalPhone, the next eight are from Babel. \textit{Train}, \textit{Dev} and \textit{Eval} columns are expressed in the number of hours available. \textit{Vow} and \textit{Con} stand for the number of distinct IPA characters used to describe vowels and consonants, respectively. Asterisk indicates tone languages.}
  \label{tab:dataset_hours}
  \centering
     \begin{tabular}{lrrrrr}
         \toprule
         Language       &  Train &   Dev &  Eval & Vow & Con \\
         \midrule
         Czech          &   24.0 &   3.3 &   3.8 & 6   & 22  \\
         French         &   22.8 &   2.0 &   2.0 & 14  & 20  \\
         Spanish        &   11.5 &   1.4 &   1.2 & 6   & 21  \\
         Mandarin$^*$   &   14.9 &   1.1 &   1.6 & 12  & 16  \\
         Thai$^*$       &   22.9 &   2.1 &   0.4 & 9   & 14  \\
         \midrule   
         Cantonese$^*$  &  126.6 &  14.1 &  17.7 & 11  & 15  \\
         Bengali        &   54.5 &   6.1 &   9.8 & 9   & 21  \\
         Vietnamese$^*$ &   78.2 &   8.6 &  10.9 & 12  & 23  \\
         Lao$^*$        &   58.7 &   6.4 &  10.5 & 10  & 17  \\
         Zulu           &   54.4 &   6.0 &  10.4 & 8   & 28  \\
         Amharic        &   38.8 &   4.2 &  11.6 & 8   & 20  \\
         Javanese       &   40.6 &   4.6 &  11.3 & 11  & 21  \\
         Georgian       &   45.3 &   5.0 &  12.3 & 5   & 20  \\
         \bottomrule
     \end{tabular}
\end{table}

\section{Results}
\label{sec:results}

We present the PTER results for the three experiments on the 13 languages in Table~\ref{tab:asr_results}. \emph{Mono} stands for monolingual, where the ASR is trained on the train set and evaluated on the eval set of a single language; \emph{Cross} stands for crosslingual, where the training sets of all the languages are used for training, except for the language that provides the eval set; and \emph{Multi} stands for multilingual, where all the training sets are combined and each language's eval set is used for scoring, i.e., the training of the language on which the system is tested is part of the overall training set.

\subsection{Multilingual improvements}

Remarkably, even though the ratio of \emph{in-language} to \emph{out-of-language} training data is low for most languages in the \textit{Multi} experiment, the improvements are consistently large. We observe 60-70\% PTER relative improvement for all GlobalPhone languages in \textit{Multi} vs \textit{Mono} scenario, and 14.1\% - 41.8\% PTER relative improvement across Babel languages. We expect that GlobalPhone data is easier to recognize due to higher signal-to-noise ratio (SNR). The best performance of European languages could be due to the higher quality of their G2P models.

Among Babel languages, Cantonese achieves the best results in the \textit{Mono} and \textit{Multi} scenarios, likely due to the highest amount of training data. In the \textit{Mono} scheme, we observe large variations in performance between languages: Bengali, Zulu and Georgian (41.2 - 43.9\% PTER) are much easier to learn than Lao, Vietnamese, Amharic and Javanese (52.1 - 58.6\% PTER). These differences cannot be explained by the amount of available training data. The addition of multilingual training data in the \textit{Multi} scheme not only consistently improves the recognition, but also flattens the performance variations - apart from Cantonese (29.8\%) and Javanese (41\%), all languages achieve similar PTER scores in the range of 32-36\%. 

\begin{table}[th]
  \caption{Phonetic token error rate (PTER) results for all 13 languages and 3 evaluation schemes. The system configurations are: Mono(lingual), Cross(lingual), Multi(lingual).}
  \label{tab:asr_results}
  \centering
  \begin{tabular}{ l l r r r }
    \toprule
    Corpus & Language & Mono & Cross & Multi \\
    \midrule
    \multirow{5}{*}{GlobalPhone} & Czech        & 26.4 & 65.8 & 8.1  \\
                                 & French       & 29.5 & 61.7 & 11   \\
                                 & Spanish      & 27.6 & 76.3 & 9.1  \\
                                 & Mandarin     & 46.3 & 85.9 & 17.2 \\
                                 & Thai         & 46   & 76.2 & 18.1 \\
    \midrule
    \multirow{8}{*}{Babel}       & Cantonese    & 36.6 & 76.9 & 29.8 \\
                                 & Bengali      & 41.2 & 81.4 & 34.6 \\
                                 & Vietnamese   & 52.3 & 75   & 35.4 \\
                                 & Lao          & 58.6 & 77.9 & 34.1 \\
                                 & Zulu         & 41.7 & 77.3 & 35.8 \\
                                 & Amharic      & 52.1 & 75.2 & 33.9 \\
                                 & Javanese     & 53   & 99.6 & 41   \\
                                 & Georgian     & 43.9 & 76.4 & 32.4 \\
    \bottomrule
  \end{tabular}
\end{table}

\subsection{Crosslingual degradations}

In the crosslingual scenario, none of the languages scored closely to the monolingual baseline. Lao, Vietnamese and Amharic PTER degraded the least (about 40\% relative), however, these languages already had the worst PTER of all monolingual baselines. French and Czech have the best cross-lingual PTER scores, suggesting some degree of similarity with one of the languages used in the training.

Among the crosslingual experiments, Javanese is an outlier with 99.6\% PTER. Analysis of the phonetic transcript output of the ASR showed that the ASR mistakenly treated Javanese as a tone language, appending tone modifiers to vowels, where there should be none. We hypothesize that the phonotactics of Javanese bias the transformer model to insert tones: like Cantonese, Thai, and Lao, Javanese syllables may begin with complex sonorant clusters, and usually end in either long vowels, nasals, or short vowel-unvoiced stop sequences.

\begin{figure}[t]
    
    \begin{subfigure}{\linewidth}
      \centering
      \includegraphics[width=\linewidth]{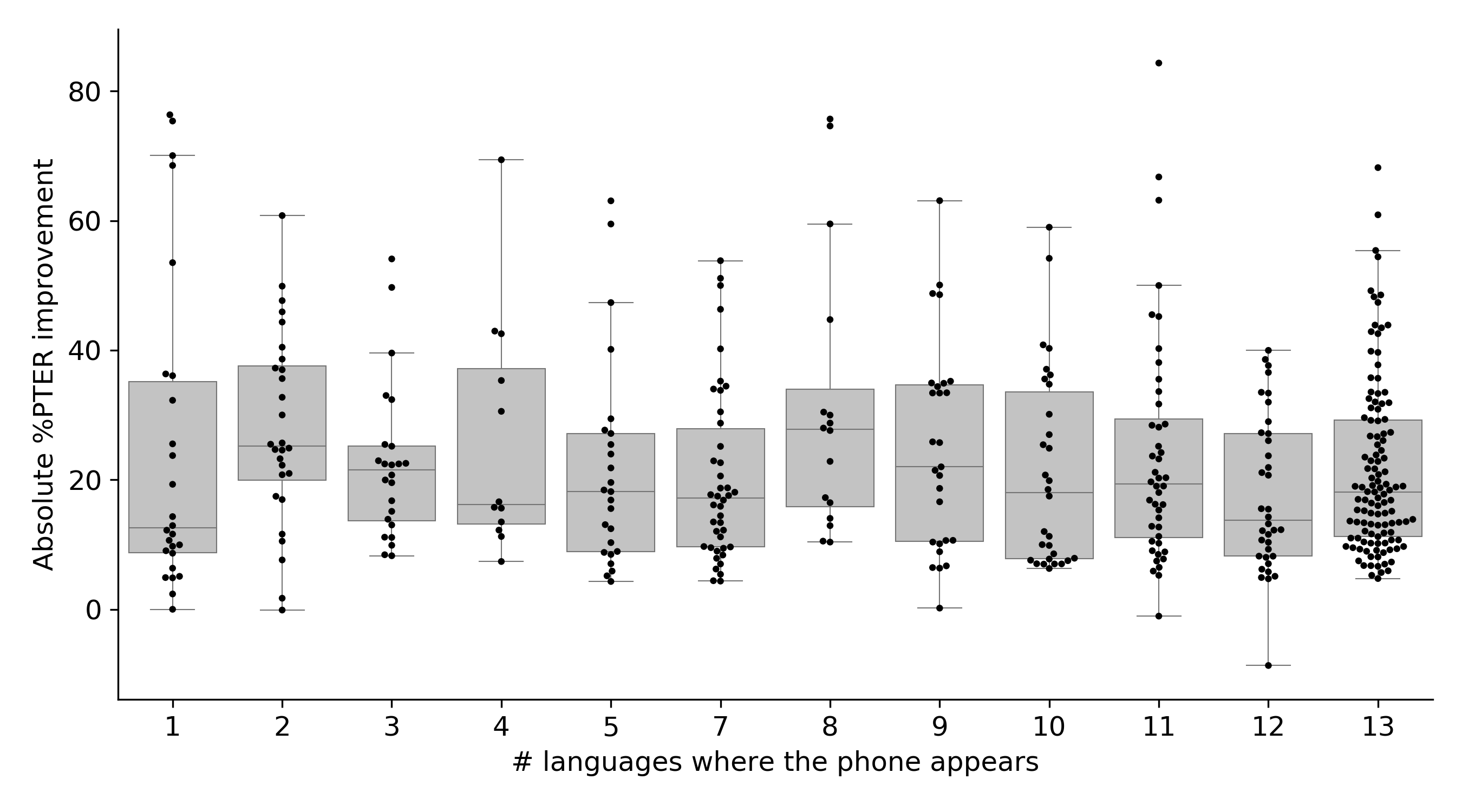}
      \caption{Improvement in the multilingual scheme vs monolingual baselines.}
      \label{fig:improvement_distributions:train_all}
    \end{subfigure}
    
    \begin{subfigure}{\linewidth}
      \centering
      \includegraphics[width=\linewidth]{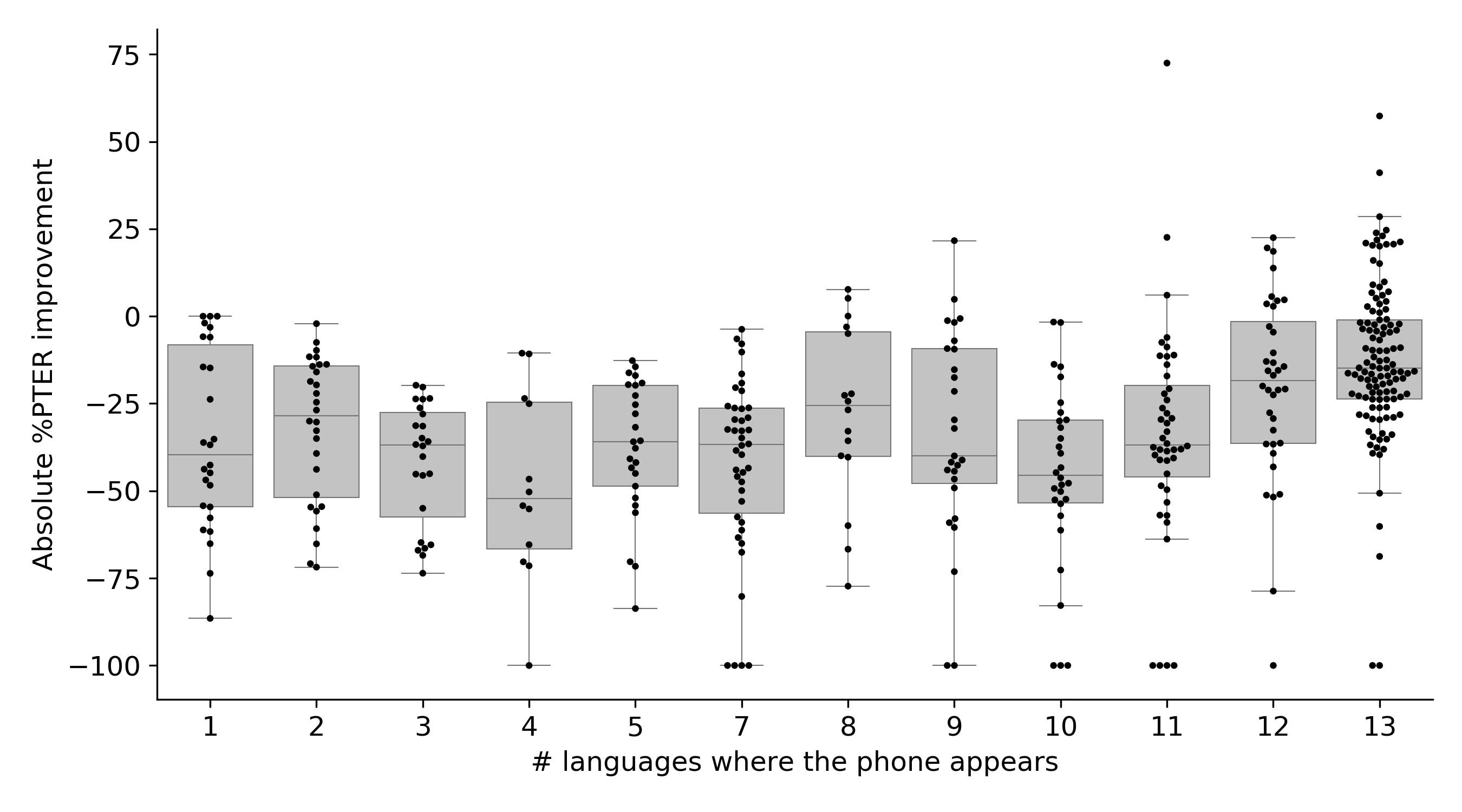}
      \caption{Degradation in the crosslingual scheme vs monolingual baselines.}
      \label{fig:improvement_distributions:oneout}
    \end{subfigure}

  \caption{Boxplot capturing the distributions of absolute PTER improvement/degradation for phones. Each column indicates the number of languages in which a given phonetic token occurs in reference transcripts. Lower numbers: the phone is more language-specific; Larger numbers: the phone is more universally occurring across languages. Each data point corresponds to a specific phone in a specific language.}
  \label{fig:improvement_distributions}
\end{figure}

\subsection{Can language-unique phones benefit from multilingual training?}

First, we seek to answer the question whether phones that exist only in a single language (language-unique phones) can still benefit from a multilingual training setup.
To that end, we show the distribution of PTER improvement for each phone, as a function of the number of languages in which the phone occurs (Fig.~\ref{fig:improvement_distributions:train_all}); the distribution of improvement for language-unique phones can be observed in the leftmost box.  Zulu has the  largest number of unique phones (\textipa{[\!g,\textbeltl,\textlyoghlig,!,||,|,\textsuperscript{H},",\textsubumlaut{ },3]}), followed by French (\textipa{[\;R,\o,\v{c},e\textrhoticity]}) and Spanish (\textipa{[B,D,\textlambda,R]}), then Bengali (\textipa{[\ae,\:r]}), Mandarin (\textipa{[\:R,\textrhookschwa]}), and Vietnamese (\textipa{[\*r,\!d]}), then Cantonese (\textipa{[5]}), Czech (\textipa{[\textraising{ }]}) and Georgian (\textipa{[q]}). Interestingly, there seem to be two groups - phones which improve slightly or do not improve at all (bottom of the distribution), and phones which gain substantially from the addition of other languages (middle and top of the distribution). The phones with top 10 improvement (31 - 74\% absolute) share very little with one another, but they each share all but one of their articulatory features with a phone that is much more common.  They include: French \textipa{[\o]} and \textipa{[e\textrhoticity]} (rounded and retroflex versions, respectively, of the more common phone~\textipa{[e]}), Vietnamese \textipa{[\!d]} and Spanish \textipa{[D]} (implosive and fricated versions, respectively, of the common phone~\textipa{[d]}), Spanish \textipa{[L]} and~\textipa{[B]} (similar to~\textipa{[l]} and~\textipa{[v]}, respectively, but with place of articulation moved one step backward or forward in the mouth), French \textipa{[\v{c}]} (identical to~\textipa{[tS]}, but annotated differently due to mismatched annotation conventions in the lexicons), and many variants of~\textipa{[r]} (Mandarin~\textipa{[\:R]}, French~\textipa{[\;R]}, and Vietnamese~\textipa{[\*r]}).  The similarity of each of these phones to another phone, in another language, suggests that these improvements could be explained by the parameter sharing inside the neural network model. This result is an important finding for the low-resource ASR community - it constitutes evidence that even though a language has unique phonetic properties, its representations can still be improved with training data from other languages.

Amongst the language-unique phones with the lowest PTER improvements, we observe two types of patterns.  First, the majority of them are classified in terms of the manner of articulation as clicks, lateral fricatives, and open-mid vowels. Second, they include all the phones unique to Zulu (\textipa{[\!g,\textbeltl,\textlyoghlig,!,||,|,\textsuperscript{H},",\textsubumlaut{ },3]}), Bengali (\textipa{[\ae,\:r]}), Cantonese (\textipa{[5]}), and Czech (\textipa{[\textraising{ }]}), and none of the phones unique to Spanish, French, Mandarin, Vietnamese, or Georgian.  It is not clear whether their low PTER is explained by their phonetic distance from the phones of other languages, or the unique properties of the languages containing them, or some combination of these factors.

\subsection{Does a phone improve more in a multilingual system when it occurs in more languages?}

The distribution of improvements in Figure~\ref{fig:improvement_distributions:train_all} does not show a clear pattern of improvement with respect to the number of languages that share a phone. The median values of improvements in each box are not necessarily increasing or decreasing as the number of languages grows. This is somewhat unexpected, as the phones shared by more languages provide more training data to the model. 

\subsection{Do phones shared by more languages transfer representations better in a crosslingual scenario?}

We also analyse the distributions of PTER degradation when comparing the crosslingual system with monolingual baselines in Figure~\ref{fig:improvement_distributions:oneout}. We observe that most of the phones' recognition rates have degraded, although this is the least for those phones shared by all languages (see the last bin in Figure~\ref{fig:improvement_distributions:oneout}). In fact, for some of these phones the PTER actually improved.  These phones are mostly consonants (improvement in 3-6 languages): \textipa{[m,n,f,l,k,s,p]}, but also vowels (2 languages each): \textipa{[u,i,a,o]}. These improvements are mostly observed in Lao (12 phones), Mandarin (10 phones), Thai and Vietnamese (4 phones each), but also Amharic, Czech and Spanish (3 phones each). 

Insertion errors make comparison between experiments difficult in some cases. For example, some improvements are observed in the language-unique phones as well - that is because the \emph{Mono} system performed a plethora of insertion errors, impossible to make in the \emph{Cross} setup. There are also outliers in the opposite direction - some phones appear only a couple of times in a given language, but they are erroneously inserted thousands of times. We clipped these outliers at -100\% PTER in the plot.

\subsection{How do manner and place of articulation affect the performance in crosslingual and multilingual schemes?}

The analysis of the overall PTER for all manners of articulation reveals that all of them have a positive mean PTER improvement when going from monolingual to multilingual scenarios. On average, flaps have the highest relative improvement (51\%) and clicks (present exclusively in Zulu), the lowest (12\%).
Regarding the place of articulation categorization, PTER improves for all categories in \emph{Multi} training versus \emph{Mono}, ranging from 29\% (retroflex category) to 44\% (uvular category). 

Analysis of PTER degradation in \emph{Cross} scheme reveals that vowels -- especially open vowels -- tend to degrade more than the other categories, excluding flaps.
Flaps have the highest degradation in \emph{Cross} models, which is caused by a large deletion rate of flaps in languages that include this category (\textipa{[\:r]} in Bengali and \textipa{[R]} in Spanish) and by insertions of flaps, especially in Czech, French and Mandarin, which do not have this category. 
A similar phenomenon occurs with dental sounds -- most of them are deleted (\textipa{[D]} in Spanish and \textipa{[|]} in Zulu), or inserted in languages that do not have dentals (again, mainly Czech, French and Mandarin).

\subsection{Are there phones with universal representations?}

\begin{figure}
  \centering
\hspace*{-0.06\linewidth}
  \includegraphics[width=1.16\linewidth]{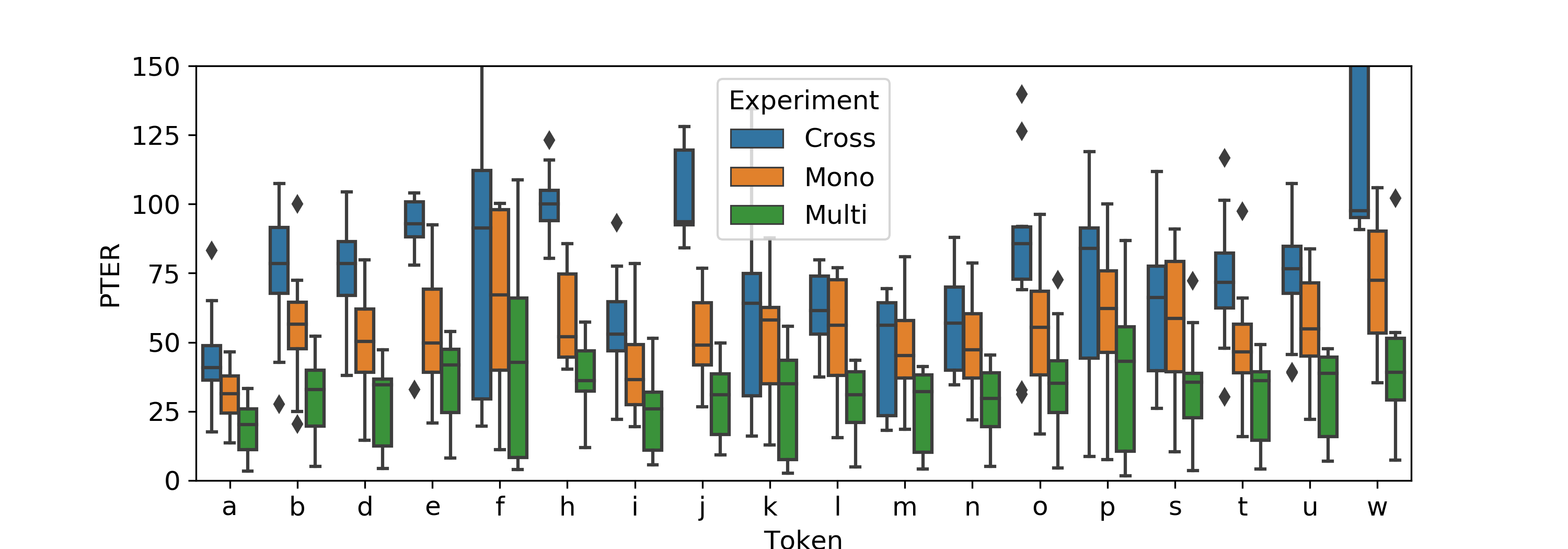}
  \caption{PTER distributions for IPA symbols ocurring in at least 11 languages.}
  \label{fig:pter_distributions_11_lang}
\end{figure}

Finally, we investigate whether any phones are recognized with a similar PTER in all experiments, which would hint at these phones having universal representations. Since the PTER variability between experiments is large, we settle at +/-25\% absolute PTER difference with regard to \emph{Mono}. With that threshold we observe that \textipa{[a]} is relatively stable, satisfying this criterion in 6 out of 12 languages (Lao does not have \textipa{[a]}). Similarly, three unvoiced plosives \textipa{[p]}, \textipa{[k]} and \textipa{[t]} (which incidentally are considered to be the most basic or universal sounds from a phonological point of view~\cite{Finegan2008}) are also stable at 5 out of 13 languages.

Approaching the question from a different direction, we examine the PTER distribution for the phones that appear in at least 11 languages in Figure~\ref{fig:pter_distributions_11_lang}. We observe that phones \textipa{[a,i,l,m,n]} have two interesting attributes: their PTER is consistently the lowest across all experiments, and the inter-quartile distance (IQD) in the \emph{Cross} condition does not seem to increase compared to \emph{Mono} or \emph{Multi} (with the exception of \textipa{[m]}, where the 25th percentile actually drops lower than in \emph{Mono}).

Although none of the results we presented allow us to definitely answer the question about the existence of universal phone representations in our model, based on the analyses concluded so far, we identify \textipa{[a,i,l,m,n,p,t,k]} as viable candidates for continued investigation.

\section{Conclusions}
\label{sec:conclusions}

In this study, we performed phone-level ASR experiments in monolingual, multilingual and crosslingual scenarios with shared IPA vocabulary and analysed the outcomes. We found major PTER improvements across all 13 languages in the multilingual setup, and stark degradation in the crosslingual systems. We observed that a crosslingual ASR system might assume that a language is tonal, even when it is not, and that the same problem does not exist when the target language data is also used in training. We have shown that all phones, even the ones unique for the target language, benefit from multilingual training. Results suggest that the benefit is not dependent on the number of languages that share the phone but rather on its similarity with phones in other languages. In contrast, phones existing in many languages do seem to degrade less in the crosslingual scenario. We did not find strong evidence of universal phone representations, even if some results suggest their existence. 

There are several questions stemming from our experiments and analyses which we will investigate in future work. Why did multilingual training work so well? There might be several factors explaining this: the phonetic variety in this particular mix of languages; innate properties of the model architecture; or the use of languages belonging to the same families. Do our conclusions generalize to languages outside of the set used in this study or to a different architecture? The analysis framework developed here can be applied to investigate that. Another interesting question is whether these major improvements in PTER would also be observed in downstream metrics such as character error rate (CER) or word error rate (WER). Such investigation must take into account language-specific vocabularies, lexicons, and language models.

\bibliographystyle{IEEEtran}

\bibliography{mybib}

\begin{thebibliography}{10}
\providecommand{\url}[1]{#1}
\csname url@samestyle\endcsname
\providecommand{\newblock}{\relax}
\providecommand{\bibinfo}[2]{#2}
\providecommand{\BIBentrySTDinterwordspacing}{\spaceskip=0pt\relax}
\providecommand{\BIBentryALTinterwordstretchfactor}{4}
\providecommand{\BIBentryALTinterwordspacing}{\spaceskip=\fontdimen2\font plus
\BIBentryALTinterwordstretchfactor\fontdimen3\font minus
  \fontdimen4\font\relax}
\providecommand{\BIBforeignlanguage}[2]{{%
\expandafter\ifx\csname l@#1\endcsname\relax
\typeout{** WARNING: IEEEtran.bst: No hyphenation pattern has been}%
\typeout{** loaded for the language `#1'. Using the pattern for}%
\typeout{** the default language instead.}%
\else
\language=\csname l@#1\endcsname
\fi
#2}}
\providecommand{\BIBdecl}{\relax}
\BIBdecl

\bibitem{samarin1984linguistic}
W.~J. Samarin, ``The linguistic world of field colonialism,'' \emph{Language in
  society}, vol.~13, no.~4, pp. 435--453, 1984.

\bibitem{helgerson1998language}
R.~Helgerson, ``Language lessons: Linguistic colonialism, linguistic
  postcolonialism, and the early modern {English} nation,'' \emph{The Yale
  Journal of Criticism}, vol.~11, no.~1, pp. 289--299, 1998.

\bibitem{pennycook2002english}
A.~Pennycook, \emph{English and the discourses of colonialism}.\hskip 1em plus
  0.5em minus 0.4em\relax Routledge, 2002.

\bibitem{Morrell2020}
E.~Morrell and J.~Rowsell, Eds., \emph{Stories from Inequity to Justice in
  Literacy Education: Confronting Digital Divides}.\hskip 1em plus 0.5em minus
  0.4em\relax New York: Routledge, 2020.

\bibitem{schultz1998multilingual}
T.~Schultz and A.~Waibel, ``Multilingual and crosslingual speech recognition,''
  in \emph{Proc. DARPA Workshop on Broadcast News Transcription and
  Understanding}, 1998, pp. 259--262.

\bibitem{loof2009cross}
J.~L{\"o}{\"o}f, C.~Gollan, and H.~Ney, ``Cross-language bootstrapping for
  unsupervised acoustic model training: Rapid development of a polish speech
  recognition system,'' in \emph{Tenth Annual Conference of the International
  Speech Communication Association}, 2009.

\bibitem{Swietojansk2012unsupervised}
P.~{Swietojanski}, A.~{Ghoshal}, and S.~{Renals}, ``Unsupervised cross-lingual
  knowledge transfer in dnn-based lvcsr,'' in \emph{2012 IEEE Spoken Language
  Technology Workshop (SLT)}, 2012, pp. 246--251.

\bibitem{knill2013investigation}
K.~M. Knill, M.~J. Gales, S.~P. Rath, P.~C. Woodland, C.~Zhang, and S.-X.
  Zhang, ``Investigation of multilingual deep neural networks for spoken term
  detection,'' in \emph{2013 IEEE Workshop on Automatic Speech Recognition and
  Understanding}.\hskip 1em plus 0.5em minus 0.4em\relax IEEE, 2013, pp.
  138--143.

\bibitem{li2020universal}
X.~Li, S.~Dalmia, J.~Li, M.~Lee, P.~Littell, J.~Yao, A.~Anastasopoulos, D.~R.
  Mortensen, G.~Neubig, A.~W. Black \emph{et~al.}, ``Universal phone
  recognition with a multilingual allophone system,'' \emph{Language Resources
  and Evaluation Conference (LREC) 2020}, 2020.

\bibitem{dalmia2018sequence}
S.~Dalmia, R.~Sanabria, F.~Metze, and A.~W. Black, ``Sequence-based
  multi-lingual low resource speech recognition,'' in \emph{2018 IEEE
  International Conference on Acoustics, Speech and Signal Processing
  (ICASSP)}.\hskip 1em plus 0.5em minus 0.4em\relax IEEE, 2018, pp. 4909--4913.

\bibitem{nagamine2015exploring}
T.~Nagamine, M.~L. Seltzer, and N.~Mesgarani, ``Exploring how deep neural
  networks form phonemic categories,'' in \emph{Sixteenth Annual Conference of
  the International Speech Communication Association}, 2015.

\bibitem{schultz2002globalphone}
T.~Schultz, ``Globalphone: a multilingual speech and text database developed at
  karlsruhe university,'' in \emph{Seventh International Conference on Spoken
  Language Processing}, 2002.

\bibitem{ghahremani2014pitch}
P.~Ghahremani, B.~BabaAli, D.~Povey, K.~Riedhammer, J.~Trmal, and S.~Khudanpur,
  ``A pitch extraction algorithm tuned for automatic speech recognition,'' in
  \emph{2014 IEEE international conference on acoustics, speech and signal
  processing (ICASSP)}.\hskip 1em plus 0.5em minus 0.4em\relax IEEE, 2014, pp.
  2494--2498.

\bibitem{kim2017joint}
S.~Kim, T.~Hori, and S.~Watanabe, ``Joint ctc-attention based end-to-end speech
  recognition using multi-task learning,'' in \emph{2017 IEEE international
  conference on acoustics, speech and signal processing (ICASSP)}.\hskip 1em
  plus 0.5em minus 0.4em\relax IEEE, 2017, pp. 4835--4839.

\bibitem{watanabe2018espnet}
S.~Watanabe, T.~Hori, S.~Karita, T.~Hayashi, J.~Nishitoba, Y.~Unno, N.-E.~Y.
  Soplin, J.~Heymann, M.~Wiesner, N.~Chen \emph{et~al.}, ``Espnet: End-to-end
  speech processing toolkit,'' \emph{Proc. Interspeech 2018}, pp. 2207--2211,
  2018.

\bibitem{watanabe2017language}
S.~Watanabe, T.~Hori, and J.~R. Hershey, ``Language independent end-to-end
  architecture for joint language identification and speech recognition,'' in
  \emph{2017 IEEE Automatic Speech Recognition and Understanding Workshop
  (ASRU)}.\hskip 1em plus 0.5em minus 0.4em\relax IEEE, 2017, pp. 265--271.

\bibitem{karita2019comparative}
S.~Karita, N.~Chen, T.~Hayashi, T.~Hori, H.~Inaguma, Z.~Jiang, M.~Someki,
  N.~E.~Y. Soplin, R.~Yamamoto, X.~Wang \emph{et~al.}, ``A comparative study on
  transformer vs rnn in speech applications,'' in \emph{IEEE Automatic Speech
  Recognition and Understanding (ASRU) Workshop 2019}.\hskip 1em plus 0.5em
  minus 0.4em\relax IEEE, 2019.

\bibitem{international1999handbook}
I.~P. Association, I.~P.~A. Staff \emph{et~al.}, \emph{Handbook of the
  International Phonetic Association: A guide to the use of the International
  Phonetic Alphabet}.\hskip 1em plus 0.5em minus 0.4em\relax Cambridge
  University Press, 1999.

\bibitem{Finegan2008}
E.~Finegan, Ed., \emph{Language: Its Structure and Use (5th edition)}.\hskip
  1em plus 0.5em minus 0.4em\relax Boston: Thomson - Wadsworth, 2008.

\end{thebibliography}

\end{document}